\documentclass[a4paper,11pt]{article}
\usepackage{pos}
\usepackage{url}
\usepackage{hyperref}
\usepackage{natbib}
\usepackage{aas_macros}
\DeclareUnicodeCharacter{03B3}{$\gamma$}
\usepackage{caption}
\usepackage{subcaption}
\usepackage{amsmath}
\usepackage{graphicx}
\usepackage{xcolor}
\usepackage{siunitx}
\title{The population of Galactic young massive star clusters in the TeV range}

\author*[a]{Rowan Batzofin}
\author[b]{Pierre Cristofari}
\author[a]{Kathrin Egberts}

\affiliation[a]{Universit\"at Potsdam, Institut für Physik und Astronomie,\\ Campus Golm, Haus 28, Karl-Liebknecht-Str. 24/25, 14476 Potsdam-Golm, Germany}

\affiliation[b]{Observatoire de Paris, PSL Research University,\\ 61 avenue de l’Observatoire, Paris, France}

\emailAdd{rowan.batzofin@uni-potsdam.de}

\abstract{Young massive star clusters (YMSCs) can produce gamma rays in the very-high-energy (VHE, E>100 GeV) range and have been proposed as sources that can accelerate cosmic rays up to PeV energies. Observations with current instruments have lead to the detection of only a few YMSCs but future instruments should significantly increase this number. However, the details of the production of the VHE emission are not well understood: What is the spectrum of accelerated particles? What is the efficiency of cosmic-ray production? What fraction of the wind luminosity is converted into the turbulent magnetic field?

To address these questions, we simulate the population of YMSCs in the gamma-ray domain, by means of Monte Carlo methods, and apply the constraints based on the subsample of  YMSCs currently detected at TeV energies. We confront our simulated populations with the catalogue of the H.E.S.S. Galactic Plane Survey and the First LHAASO Catalogue of Gamma-Ray Sources, allowing us to investigate crucial aspects of particle acceleration at YMSCs.
}

\ConferenceLogo{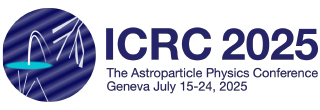}

\FullConference{39th International Cosmic Ray Conference (ICRC2025)\\
 15–24 July 2025\\
Geneva, Switzerland\\}

\begin{document}
\maketitle
\section{Introduction}
Cosmic rays (CRs) were discovered in the 1910s thanks to the pioneering work of Domenico Pacini and Victor Hess and have been extensively studied ever since  \citep{Hess:1912srp}. The most fundamental question of CR physics, surprisingly, remains unanswered: Where are they produced? Of particular interest is the origin of cosmic rays that reach PeV energies. Among the popular hypotheses is that they can be accelerated by young massive star clusters (YMSCs), via the first-order Fermi mechanism called diffusive shock acceleration (DSA) \citep{Blandford_1978, Morlino_2021}. DSA can efficiently energise protons (and ions), as well as electrons. Subsequently, the accelerated protons and electrons can interact with the interstellar medium (ISM) to produce gamma-rays in the very-high-energy (VHE) domain, mainly through two mechanisms: 1) the production/decay of neutral pions, in the collision of accelerated protons with hadrons of the ISM (hadronic mechanism), and 2) the inverse Compton scattering of accelerated electrons on soft photons: cosmic microwave background (CMB), infrared, optical (leptonic mechanism). For YMSCs the dominant process is thought to be hadronic.

 H.E.S.S. (the High Energy Stereoscopic System), an array of imaging atmospheric Cherenkov telescopes located in the Khomas highlands in Namibia, performed a systematic survey of the Galactic plane. The H.E.S.S. Galactic Plane Survey (HGPS) covers the Galactic plane from longitudes of $250^{\circ}$ to $65^{\circ}$ and from latitudes of $-3^{\circ}$ to $3^{\circ}$. This systematic survey found 78 sources in the TeV domain. Among these sources is the stellar cluster Westerlund 1 \citep{westerlund_1}.
LHAASO (Large High Altitude Air Shower Observatory), a multi-purpose ground-based extensive air shower array located in Daocheng County, Sichuan Province, China, released a catalogue of very-high-energy and ultra-high-energy gamma-ray sources \citep{LHAASO_catalogue}. The First LHAASO Catalogue of Gamma-Ray Sources covers the sky above 1 TeV, it covers declinations from $-20^{\circ}$ to $80^{\circ}$. There are 90 sources in the catalogue, 43 of which have an energy above 100 TeV. Among the gamma-ray emission observed by LHAASO is emission coming from the super bubble in the direction of the star forming region Cygnus-X \citep{LHAASO_cygnus}.

We intend to simulate the population of TeV emitting YMSCs and confront it with the population detected in the HGPS and the First LHAASO Catalogue of Gamma-Ray Sources. For this purpose, we simulate populations of Galactic YMSCs using a Monte Carlo method. We simulate the positions, masses, and wind luminosities of the YMSCs and calculate the gamma-ray emission from the YMSCs at the current time. We constrain the simulations with measurements from the HGPS and the First LHAASO Catalogue of Gamma-Ray Sources by identifying the detectable sources in the simulations and comparing them with the detected sample of YMSCs. For the HGPS the detectability is based on the longitude, latitude, and angular-extension dependent HGPS sensitivity. The inclusion of the complex dependencies of the HGPS sensitivity is crucial for drawing conclusions on the YMSC population based on this data comparison. For LHAASO detectability we consider the two instruments (Water Cherenkov Detector Array (WCDA) and Kilometre Squared Array (KM2A)) separately and take into account the declination and extent of the simulated sources.  
\section{YMSC population model} \label{section:pop_model}

Similarly to other population simulations \citep{1st_paper}, in order to simulate the population of Galactic YMSCs emitting in the TeV range, we need: a) a model for particle acceleration at the YMSC and the subsequent gamma--ray emission (from hadronic interactions of accelerated particles with the surrounding medium), which needs to be computationally inexpensive; b) a description of the spatial distribution of YMSCs in the Galaxy.

Following the work of \citep{Menchiari_2023, Menchiari_2024}, we assume that the collective wind of a YMSC can be described in the same way as the wind of a single star with the total mass of the cluster in the centre, the mass-loss rate as the sum of all the mass-loss rates in the cluster, and the wind luminosity as the sum of all the wind luminosities in the cluster. The evolution of YMSCs follows the approach of \citep{Weaver_1977}, the bubble radius ($R_{\rm b}$) and the radius of the termination shock ($R_{\rm ts}$) are calculated as in \cite{Menchiari_2023} and \citep{Menchiari_2024}, taking into account the wind luminosity, wind mass-loss rate, density, wind velocity, and age of the cluster. Looking only at the total flux of the cluster and not the individual contributions of the sources in the cluster, we estimate the gamma-ray emission above 1 TeV from the distribution of CRs accelerated in the winds of the YMSCs. We assume that some fraction, $\eta_{\rm CR}$, of the wind luminosity is converted into CRs and that some fraction, $\eta_{b}$, of the wind luminosity is converted into magnetic flux. The maximum energy of the accelerated CRs is estimated using the intrinsic properties of the YMSCs by equating the diffusion length to the size of the TS: $\frac{D_1 (E_{\rm max})}{u1}=R_{\rm TS}$, where $u1$ is the upstream shock velocity. The gamma-ray emission of YMSCs is calculated from the spectrum of protons assuming a spectral index, $\alpha$, and that the target medium is the same as the density of the ambient medium in which the bubble expands (set to $10m_p$ $\rm cm^{-3}$).

The simulated YMSCs are drawn relying on a Monte Carlo approach: the age and total masses of the clusters are drawn following the distribution of the ages and masses in the Global survey of star clusters in the Milky Way \citep{Piskunov_2018}, respectively. A mass for each star is drawn from the distribution of star masses \citep{kroupa_mnras_322_2001}, from which the wind luminosity and mass-loss rate are estimated. The clusters are distributed according to the distribution of ISM following the model of \citep{Steiman-Cameron_2010}. In this work, we are interested in clusters that are: $<10$ Myrs old and have total mass $> 1000 \: M_\odot$. The total number of such clusters can be estimated using: $N_{SC} = D \int_{M_{min}}^{M_{max}} \int_{t_{min}}^{t_{max}} \int_{0}^{R_{MW}} r \xi_{SC} \left(M,t,r\right)dM \: dt \: dr \: d\theta$, where $R_{MW}$ is the radius of the Milky Way, $\xi_{SC}$ is the cluster distribution function and D is a normalisation constant. We use $\xi_{SC}$ in the following form: $\xi_{SC}\left(M,t,r\right) = f(M) \nu(t) \rho(r,\theta)$, where $f(M)$ is the cluster mass function \cite{Piskunov_2018}, $\nu(t)$ is the cluster formation rate \citep{Piskunov_2018} and $\rho(r,\theta)$ is the spatial distribution \citep{Steiman-Cameron_2010}. To find the normalisation constant we consider all the clusters in the completeness region (within 1.8 kpc) of the Global survey of star clusters (1581 clusters). We calculate the number of expected clusters ($N_{SC}$) of all ages and masses within 1.8 kpc of the Sun and solve for D such that the number of clusters is 1581. We can then estimate the number of YMSCs we expect, which we find to be $\sim 16$.

\section{Confronting our simulated populations to available experimental data} \label{section:simulating_populations}

The parameters of interest in our simulated YMSCs are: $\alpha$ : the spectral index of the accelerated particles, $\eta_{CR}$ : the fraction of the wind luminosity converted into CRs, $\eta_b$ : the fraction of the wind luminosity converted into turbulent magnetic field, and the diffusion regime. We typically compute 100 realisations for a given set of parameters, to ensure stability of our results.  Note that the positions of simulated YMSCs change for each of the 100 realisations for a given set of properties but these 100 different position sets are constant when changing the population properties. This allows us to closely examine how each of the properties affects the population.

We consider two different Galactic surveys, the HGPS and the First LHAASO Catalogue of Gamma-Ray Sources. For each source, using the gamma-ray luminosity, size and position we are able to determine whether it would have been detected in the HGPS and the First LHAASO Catalogue of Gamma-Ray Sources. To account for the selection bias in the H.E.S.S. catalogue, we follow the work in \citep{Steppa_2020} and in particular account for the fact that the HGPS sensitivity varies with position and source extent. For a simulated source to be detectable, it must exceed the threshold of $5\sigma$ above the background.

\begin{figure}[h]
    \centering
    \includegraphics[width = \textwidth]{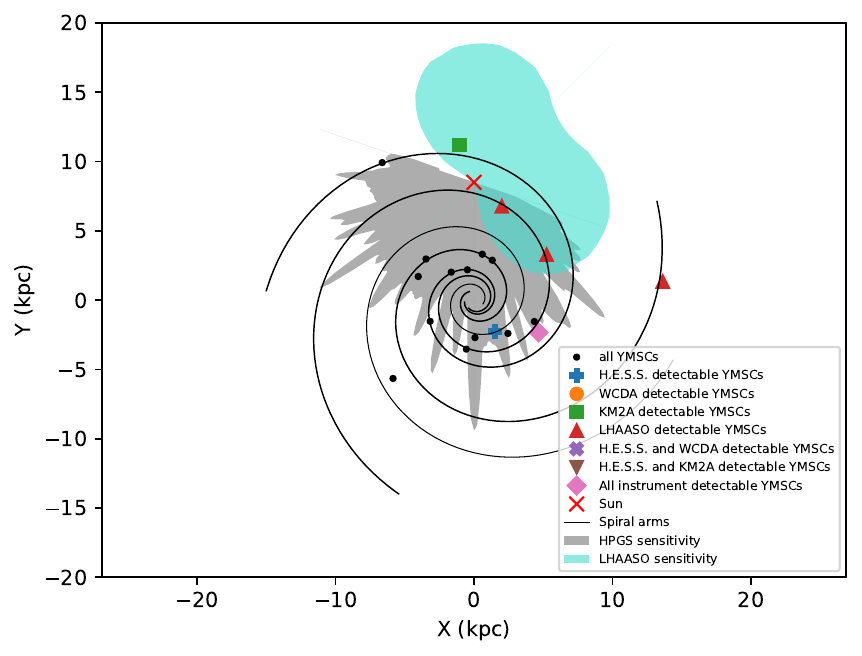}
    \caption{Example of one realisation of a simulation of the population of Galactic YMSCs. The grey shaded region is the HGPS detectability range for point like sources with a luminosity of $5\times 10^{33}$ photons $\text{s}^{-1}$ (equivalent to a differential flux at 1 TeV of $\sim 4 \times 10^{-11} \rm TeV^{-1} cm^{-2} s^{-1}$ at a distance of 1 kpc). The turquoise shaded region is the LHAASO detectability range for point like sources with a luminosity of $1\times 10^{32}$ photons $\text{s}^{-1}$ and $3\times 10^{27}$ photons $\text{s}^{-1}$ (equivalent to a differential flux of $\sim 3 \times 10^{-13} \rm TeV^{-1} cm^{-2} s^{-1}$ and $\sim 5 \times 10^{-19} \rm TeV^{-1} cm^{-2} s^{-1}$ at 3 TeV and 50 TeV respectively at a distance of 1 kpc) for the WCDA and KM2A respectively. The black dots mark the simulated sources which are not detectable by any of the instruments, the blue plusses mark the simulated sources only detectable by H.E.S.S., the green squares mark sources only detectable by the KM2A, the red triangles mark sources detectable by both LHAASO instruments and the pink diamonds mark sources detectable by all the instruments. 26\% of the YMSCs are detectable, by at least one instrument, in this realisation. The Sun is marked by the red cross and the spiral arms of the Milky Way are shown in black.}
    \label{fig:single_realisation}
\end{figure}

The limited field of view of H.E.S.S. combined with the applied technique of deriving background measurements means that sources $\gtrsim 1^\circ$ are not detectable. The detections for LHAASO are split by detector (WCDA and KM2A), therefore, we need a function for each detector to determine detectability. The sensitivity of each detector as a function of declination can be seen in \citep{LHAASO_catalogue} figure 5. We consider the position, integrated flux (above 3 TeV for the WCDA and above 50 TeV for the KM2A), and extent of the source when determining its detectability. In Fig. \ref{fig:single_realisation} the HGPS detectability range for point like sources with a luminosity of $5 \times 10^{33} \text{photons s}^{-1}$ is displayed, demonstrating the large variations in sensitivity as a function of Galactic longitude. This variation in sensitivity shows why it is not sufficient to have a cutoff in the integrated flux to determine if a source is detectable, and thus why we need to include the sensitivity of H.E.S.S. when testing each simulated source. Also seen in Fig.~\ref{fig:single_realisation} are the different combinations of instruments detecting the simulated sources, the realisation shown has the following parameters: $\alpha=4.5$, $\eta_{\rm CR} = 0.1$, $\eta_b = 0.1$, and using the Bohm diffusion regime.

Because many of the sources in the catalogues are unidentified, and at least some of those could be YMSCs, we created upper and lower limits for each of the catalogues we compare with. When comparing with the HGPS, the lower limit is the known YMSC, Westerlund 1 \citep{westerlund_1}, and the upper limit is the lower limit plus the unidentified sources associated by position with known stellar clusters (5). For the LHAASO detectors we do the same thing, the lower limit (for both the WCDA and the KM2A) is the Cygnus region \citep{LHAASO_cygnus} and the upper limits are the lower limit plus the sources associated by position with known stellar clusters, 12 for the WCDA and 9 for the KM2A. It should be noted that the upper limits for the LHAASO detectors are less important than the lower limit since almost all of our simulated populations that do not match the data have no detectable sources, as opposed to too many detectable sources. Additionally, we have one more criterion: there should be at least two separate YMSCs detectable, since it is possible that one simulated YMSC can be detectable by all three instruments, but we have detected at least two in the Galaxy.

\section{Results} \label{section:results}
\begin{figure*}
    \centering
    \includegraphics[width = \linewidth]{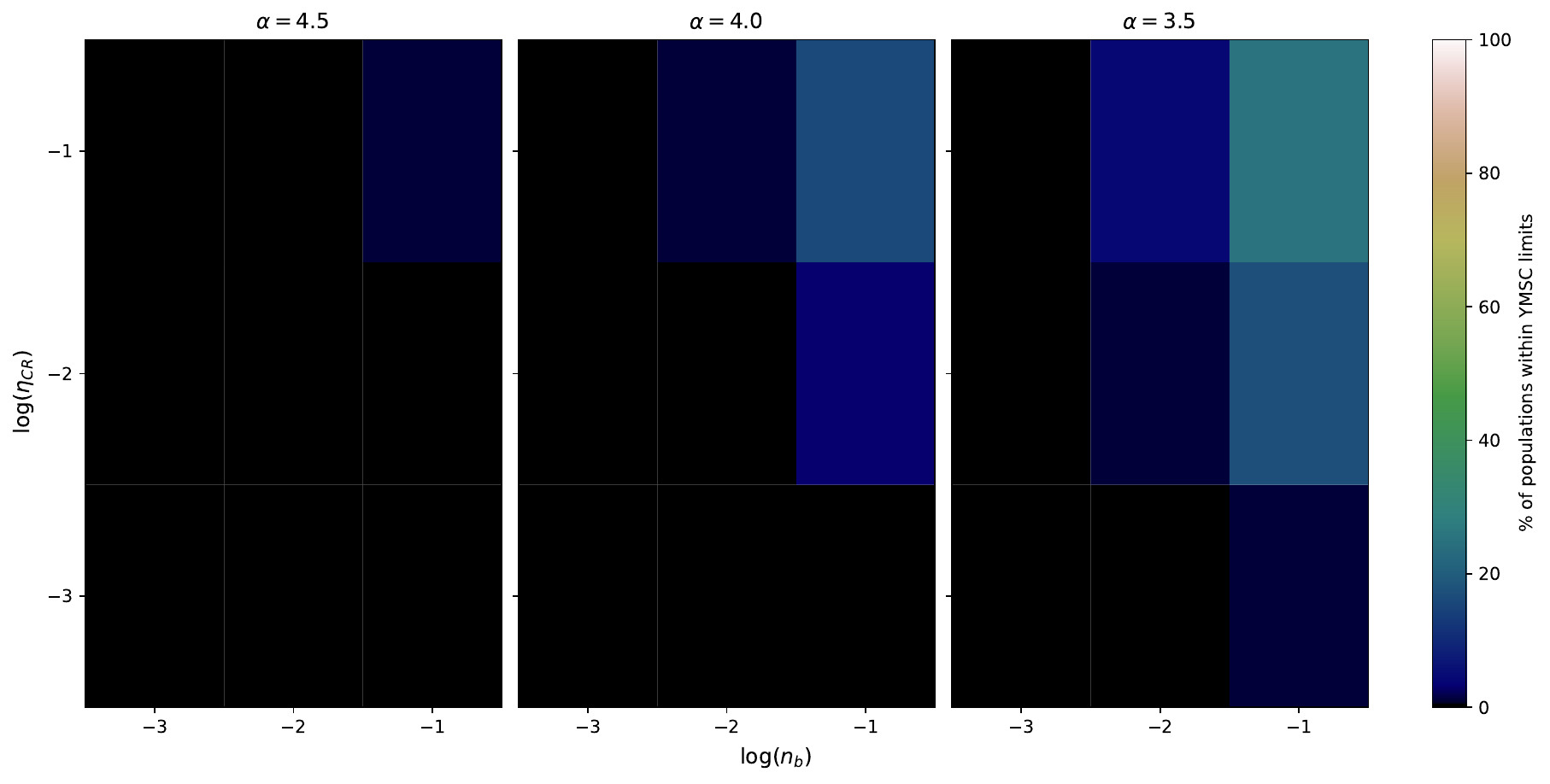}
    \caption{2D histogram plots showing the percentage of populations are in agreement with the observational data, i.e. have at least 2 detectable YMSCs and are within the upper and lower limits for all three detectors (H.E.S.S., WCDA and KM2A). These populations use the Kraichnan diffusion regime. The x-axis shows the fraction of the wind luminosity converted into cosmic rays, the y-axis shows the fraction of the wind luminosity converted into magnetic field. Moving from left to right shows a hardening spectral index.}
    \label{fig:Kraichnan}
\end{figure*}

\begin{figure*}
    \centering
    \includegraphics[width = \linewidth]{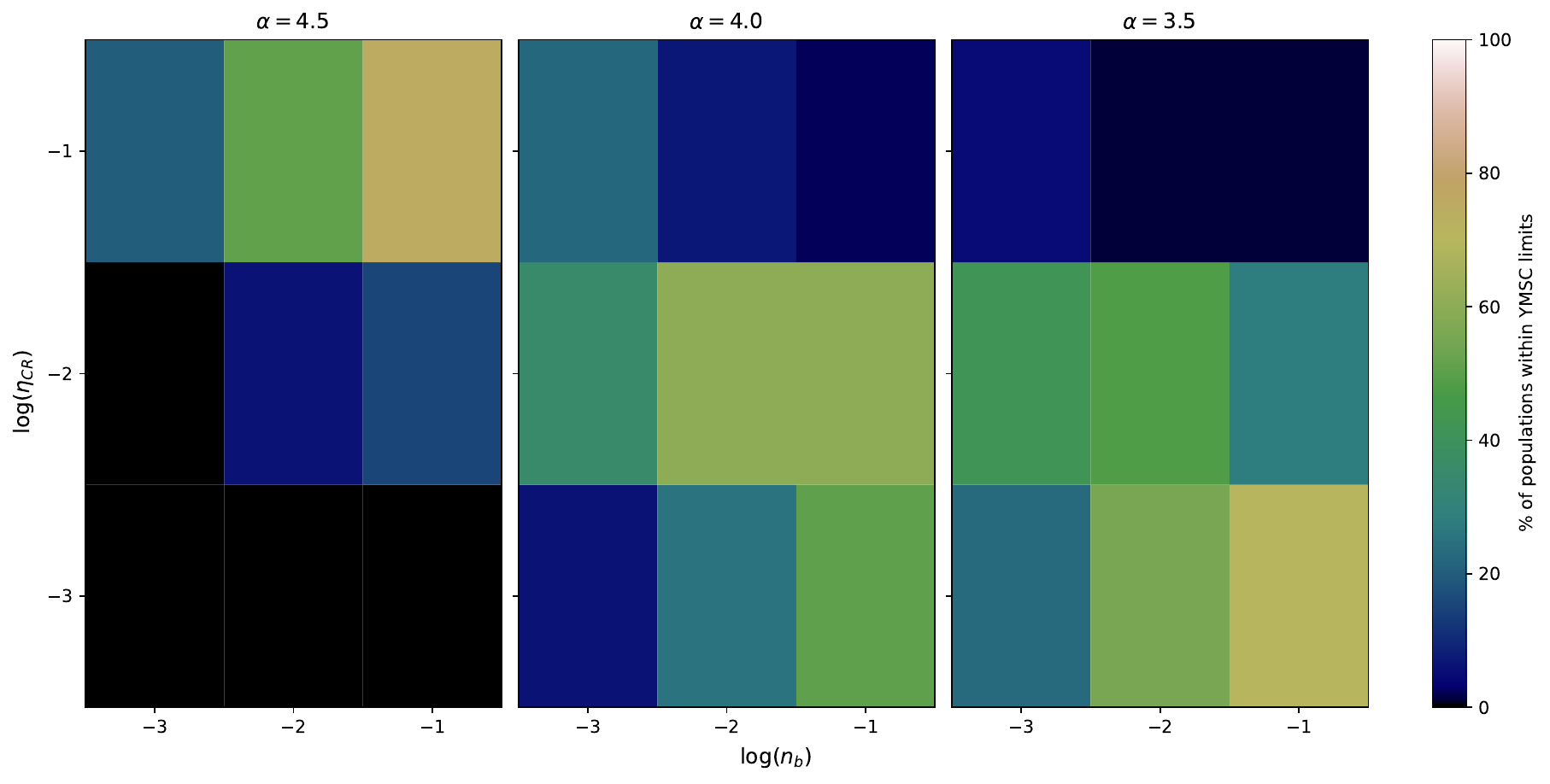}
    \caption{Same as Fig.~\ref{fig:Kraichnan} except with Bohm diffusion.}
    \label{fig:Bohm}
\end{figure*}

We confronted our synthetic populations with both the HGPS and the First LHAASO Catalogue of Gamma-Ray Sources. We performed a systematic search of our parameter space: $3.5 \leq \alpha \leq 4.5$, $10^{-3} \leq \eta_{\rm CR} \leq 0.1$, $10^{-3} \leq \eta_{b} \leq 0.1$, and the following diffusion regimes: Kolmogorov, Kraichnan, and Bohm. We confront our synthetic populations to the HGPS and the First LHAASO Catalogue of Gamma-Ray Sources, and investigate the situations when our simulated populations are found in agreement with the experimental data, by requiring that the detectable sources in our populations fit into the limits for each detector (H.E.S.S., WCDA and KM2A) and that there are at least 2 detectable sources (by any combinations of instruments) in our synthetic populations. The results of this search can be seen in Figs.~\ref{fig:Kraichnan} and \ref{fig:Bohm}. As you decrease $\alpha$ or increase $\eta_{CR}$ and $\eta_b$ the integrated flux of the YMSCs increases and so does the maximum energy particles can be accelerated to.

We found that no populations matched the data when using the Kolmogorov diffusion regime. None of the populations had enough detectable simulated YMSCs to match the data. When considering only the HGPS data, we found one parameter set ($\alpha = 3.5$, $\eta_{CR} = 0.1$ and $\eta_b = 0.1$) with 17\% of the populations within the limits. When considering only the WCDA data, we found that the same parameter set (also the best matching) had 8\% of populations within our limits. When confronting the populations with only the KM2A data, we found that the best-matching parameter sets (the previous parameter set included) had only 1\% of populations within our limits. Fig.~\ref{fig:Kraichnan} shows the parameter search for the parameter sets with Kraichnan diffusion. The Kraichnan diffusion regime accelerates particles to higher energies than the Kolmogorov regime and we have 25\% of populations in agreement with all the data for the best parameter set ($\alpha = 3.5$, $\eta_{CR} = 0.1$ and $\eta_b = 0.1$), this set can be seen in the top right part of the right most panel in Fig.~\ref{fig:Kraichnan}. Again, the most limiting instruments for the Kraichnan case are the LHAASO WCDA and KM2A detectors. When confronting our populations only with the HGPS we find that for one parameter set ($\alpha = 3.5$, $\eta_{CR} = 0.1$ and $\eta_b = 0.1$) 83\% of the populations are in agreement. This drops to 77\% and 47\% for the WCDA and KM2A detectors, respectively.

Fig.~\ref{fig:Bohm} shows the parameter search for the parameter sets with Bohm diffusion. The Bohm diffusion regime accelerates particles to higher energies than both the Kolmogorov and Kraichnan regimes. 75\% of the populations for the best parameter set ($\alpha = 4.5$, $\eta_{CR} = 0.1$ and $\eta_b = 0.1$) are in agreement with all the data. With the Bohm diffusion regime we are no longer only rejecting populations  because we are not detecting enough simulated YMSCs, but also because we are detecting too many simulated YMSCs. We found that our best parameter set ($\alpha = 4.5$, $\eta_{CR} = 0.1$ and $\eta_b = 0.1$) has 87\% of the populations in agreement when only considering the HGPS. Similarly, 99\% and 98\% of populations are in agreement with WCDA and KM2A, respectively. In fact, for many different parameter sets >90\% of populations are in agreement with the LHAASO data. This clearly shows that combining the information from the different detectors helps to refine our results.

\section{Conclusion} \label{section:conclusion}

Relying on a Monte Carlo approach, we simulated populations of the Galactic YMSCs in the TeV range (built on a physically motivated model of the gamma--ray emission from accelerated protons and electrons at YMSC shocks) and confronted the results of the simulated populations to available data from the systematic survey of the Galactic plane performed by H.E.S.S. and of the First LHAASO Catalogue of Gamma-Ray Sources. We performed a systematic exploration of the parameter space defined by: $\alpha$, the power-law spectral index of the accelerated particles at the YMSC shocks; $\eta_{CR}$, the fraction of the wind luminosity converted into CRs; $\eta_{b}$, the fraction of the wind luminosity converted into a turbulent magnetic field; and considering different diffusion regimes.
Our conclusions are: 1. The systematic exploration of the parameter space allows us to look for the region of the parameter space producing the most realisations within the limits of HGPS sample and the First LHAASO Catalogue of Gamma-Ray Sources. In addition to the requirement on the total number of sources detectable by each instrument, including the total number detected in the population by any instrument, we can further reduce the allowed parameter space. One possible solution rendering $\sim 75$\% populations in the combined limits of the HGPS and First LHAASO Catalogue of Gamma-Ray Sources is: $\alpha = 4.5$, $\eta_{CR} = 0.1$, $n_b = 0.1$, and assuming the Bohm diffusion regime. 2. Some regions of our parameter space seem to be disfavoured: Kolmogorov diffusion is completely disfavoured. Kraichnan diffusion is mostly disfavoured with only a few parameter sets producing any populations in agreement with the data, and at most 25\%. When $\alpha = 4.5$, $\eta_{\rm CR} \leq 10^{-2}$ is disfavoured, whereas for $\alpha \geq 4.0$, $\eta_{\rm CR} > 10^{-2}$ is disfavoured.

A finer binning in the exploration of our parameters would likely lead to some more populations that are in better agreement with the experimental data. Moreover, an increase in the sample of the firm YMSC detections would help our approach since at this moment there are very few (2). This increased number of detections is clearly expected with next-generation instruments such as CTAO~ \citep{cristofari2017,scienceCTA,CTA_galactic} or the LHAASO wide field air Cherenkov telescopes array~ \citep{LHAASO_IACT}. Including SNR shocks or Wolf-Rayet stars in the modelling of the YMSCs could produce particle acceleration to higher energies than our current model and produce more detectable simulated YMSCs.


\bibliography{Population_Galactic_YMSCs}
\end{document}